\documentstyle[prl,aps,epsfig,preprint]{revtex}
\newcommand{\bm}{\bibitem}

\def\be {\begin{equation}}
\def\ee {\end{equation}}
\def\bea {\begin{eqnarray}}
\def\eea {\end{eqnarray}}
\def\nn {\nonumber}

\begin{document}
\title{Can collisional energy loss explain  
nuclear suppression factor for light hadrons ? }
\author{Jan-e Alam$^a$,  
Pradip  Roy$^b$ and Abhee K. Dutt-Mazumder$^b$  } 
\address{a) Variable Energy Cyclotron Centre, 1/AF Bidhannagar, Kolkata, India}
\address{b) Saha Institute of Nuclear Physics, 1/AF Bidhannagar, Kolkata, India}
\maketitle

\vspace{0.5cm}
\begin{abstract}
We argue that in the measured $p_T$ domain of RHIC, collisional rather than
the radiative energy loss is the dominant mechanism for jet quenching. 
Accordingly we calculate nuclear suppression factor for light hadrons
by taking only the elastic energy loss in sharp contrast with the
previous calculations where only the radiative loss are considered.
\\[0.1 cm]
{PACS numbers: 12.38.Mh, 24.85.+p, 25.75.-q,13.87.Fh }
\end{abstract}
\vspace{0.2cm}

Jet quenching is one of the most promising tools to extract the initial
parton density produced in high energy heavy ion collisions.
This is related to the final state energy loss of the leading partons 
\cite{bjorken,plumer,gyulassyreview} 
causing depopulation of hadrons at high transverse momentum
(see~\cite{npa757} for experimental
results). The  suppressions of high $p_T$ hadrons and  unbalanced 
back-to-back azimuthal correlations of the dijet events
measured at Relativistic Heavy Ion 
Collider (RHIC) provide experimental evidence in
support of the quenching. Based on the calculations performed
by several 
authors~\cite{rad1,rad2,zakharov} 
the detailed theory of `jet tomography' 
was developed
by Gyulassy {\em et al.}\cite{gyulassyreview} considering only the energy 
loss due to induced bremsstrahlung radiation. The observed nuclear suppression
of light hadrons ($\pi, \eta$) in $Au+Au$ collisions at
$\sqrt{s}=62-200$ AGeV at 
RHIC could
be accounted for in these models. In all these analyses the collisional loss  
was ignored~\cite{wang,salgado}. 
The non-photonic single electron spectrum from heavy meson decays measured by
PHENIX Collaboration~\cite{raacharm} put this assumption in question. 
No realistic parameter set can
explain this data using the radiative energy loss based jet tomography
model which either requires violation of bulk entropy bounds or 
non-perturbatively large $\alpha_s$ of the theory \cite{wick06}, or equivalently
one requires excessive transport co-efficient $\hat{q}_{\rm eff}= 14$ GeV$^2$/fm
\cite{armesto05}.

The importance of collisional loss in the context of RHIC was first 
discussed by the present authors \cite{abhee05,roy06}. It is shown 
in ref.\cite{abhee05} that there exists an energy range where collisional loss 
is as important as or even greater than its radiative counter part, hence
cannot be neglected in any realistic model of jet quenching. 
Recently this is also noted in ref.\cite{mustafa,wick06,djordjevic06,speigne}.
It is similar to the passage of charged particles through material
medium where the ionization loss is know to be the dominant mechanism 
at lower energies while at higher energies bremsstrahlung takes over.
There exists a critical energy $E_c$ at which they contribute equally {\rm i.e.} 
$(dE/dx)_{rad}=(dE/dx)_{coll}$ at $E = E_c$. For example, for an electron 
(proton) 
traversing copper target $E_c \sim 25 $ MeV (1 GeV)~\cite{leo}. Note that for 
heavier particle $E_c$ is higher. This indicates that for the heavy quark 
collisional loss may be more important than the radiative loss at
intermediate energies. 
In ref.\cite{abhee05} we have calculated $E_c$ for light partons under RHIC
conditions.

In this light, we, in the present work would like to address if 
the omission of collisional loss at RHIC is justified or not. 
We argue that, whether the collisional or radiative loss is the main mechanism
is a $p_T$ dependent question. It also depends
on the energies of the colliding system and expected to be different
for RHIC and Large Hadron Collider (LHC). In contrast to the previous
works, we,  therefore,  calculate nuclear suppression factor $(R_{AA})$
for pions considering only the collision energy loss. At the end we shall
show that there exists a $p_T$ window where this is reasonable assumption
contrary to the commonly held view that collisional loss (for light partons) 
can be ignored altogether.

The neutral 
pion production~\cite{nuclex06} (for  charged hadrons see ~\cite{STAR})
at RHIC in the $p_T$ window $\sim 1-13$ GeV,  is found to be suppressed
compared to the binary scaled $p$-$p$ estimation~\cite{npa757}. 
This is attributed to the
final state energy loss of the partons while passing through the plasma
before fragmenting into hadrons   
\cite{prl01,prl02,prl02p}. The energy loss in the standard perturbative 
calculations can
be incorporated by modifying the fragmentation function. This is accomplished
by replacing fractional momentum $z$ carried by the hadrons with  
$z^*=z/(1-\Delta z)$
in the argument of the fragmentation function, $D(z,Q^2)$, where 
$\Delta z = \Delta E/E$. This implementation assumes that all the partons 
suffer equal amount of energy loss which is questionable as argued in 
ref.\cite{jeon,baierjhep}. We, therefore, take a different approach where the 
initial spectra is evolved dynamically by using Fokker Planck (FP)
equation.
FP equation  can be derived from Boltzmann equation 
if the collisions are dominated by the small
angle scattering involving soft momentum exchange
\cite{roy06,alamprl94,svetitsky,moore05,ducati,rajuprc01,rapp}. 
For an expanding plasma, 
FP equation takes the following form:
\bea
\left (\frac{\partial}{\partial t}
-\frac{p_\parallel}{t}\frac{\partial}{\partial p_\parallel}\right )f({\bf p},t)
&&=\frac{\partial}{\partial p_i}[p_i\eta f({\bf p},t)]\nn\\ 
&&+\frac{1}{2}
\frac{\partial^2}{\partial p_\parallel^2 }[{B_\parallel}({\bf p})f({\bf p},t)]
\nn\\
&&+\frac{1}{2}
\frac{\partial^2}{\partial p_\perp^2}[{B_\perp}f({\bf p},t)],
\label{fpexp}
\eea
where the second term on the left hand side arises due to expansion. 
Bjorken hydrodynamical model~\cite{bjorken1983}  
has been used here for space time evolution.
In Eq.~(\ref{fpexp})  
$\eta$ denotes drag coefficient which is related to the energy loss or
the `stopping power' of the plasma, $\eta=(1/E)dE/dx$.
$B_\parallel$, $B_\perp$ denote diffusion constants along parallel and
perpendicular directions of the propagating parton representing rate
of longitudinal and transverse broadening (variance) {\rm i.e.} 
$B_\parallel=d\langle(\Delta p_\parallel)^2\rangle/dt$,
$B_\perp=d\langle(\Delta p_\perp)^2\rangle/dt$. These transport 
coefficients can be calculated from the following expressions: 

\bea
\frac{dE}{dx}&&=\frac{\nu} {(2\pi)^5}
\int\frac{d^3kd^3qd\omega}{2k2k^\prime 2p 2p^\prime}
\delta(\omega-{\bf v_p\cdot q})
\delta(\omega-{\bf v_k\cdot q})\nn\\
&&\langle {\cal M} \rangle_{t\rightarrow 0}^2
f(|{\bf k}|)\left[1+f(|{\bf k}+{\bf q}|)\right]
\omega.
\eea

\bea
B_{\perp,\parallel}&&=\frac{\nu }{(2\pi)^5}
\int\frac{d^3kd^3qd\omega}{2k2k^\prime 2p 2p^\prime}
\delta(\omega-{\bf v_p\cdot q})
\delta(\omega-{\bf v_k\cdot q})\nn\\
&&\langle {\cal M} \rangle_{t\rightarrow 0}^2
f(|{\bf k}|)\left[1+f(|{\bf k}+{\bf q}|)\right]
q_{\perp,\parallel}^2.
\eea
In the above equations the small angle limit has been taken 
to write the arguments of
the delta functions \cite{abhee05,roy06}.

The  matrix elements include
diagrams involving exchange of massless gluons which render
$\eta$ and $B_{\parallel,\perp}$ infrared divergent. Such
divergences can naturally be cured by using the hard thermal
loop (HTL)~\cite{pisarski} corrected propagator for the gluons 
as discussed below.  
We work in the coulomb gauge where the gluon propagator for the
transverse and longitudinal modes are
denoted by  $D_{00}=\Delta_{\parallel}$ and
$D_{ij}=(\delta_{ij}-q^i q^j/q^2)\Delta_{\perp}$ 
with~\cite{lebellac}:
\bea
\Delta_{\parallel}(q_0,q)^{-1}=q^2-\frac{3}{2}\omega_p^2
\left [
\frac{q_0}{q}ln\frac{q_0+q}{q_0-q}-2
\right ]
\eea

\bea
\Delta_{\perp}(q_0,q)^{-1}=q_0^2-q^2+\frac{3}{2}\omega_p^2
\left [\frac{q_0(q_0^2-q^2)}{2q^3}
ln\frac{q_0+q}{q_0-q}-\frac{q_0^2}{q^2}
\right ]
\eea

The HTL modified  matrix element in the limit of small angle scattering
takes the following form \cite{abhee05,roy06} for all the partonic
processes having dominant small angle contributions like $qg\rightarrow qg$,
$qq\rightarrow qq$ etc. :
\bea
|{\cal{M}}|^2&=& g^4 C_{R} 16 (E_p E_k)^2
\vert \Delta_{\parallel}(q_0,q)  \nn\\
&+& ({\bf{v_p}}\times {\hat{q}}).({\bf{v_k}} \times \hat{q})
\Delta_{\perp}(q_0,q)\vert^2
\eea
where $C_{R}$ is the appropriate color factor. 
With the screened interaction,  
the drag and diffusion constants can be calculated along the line of 
ref.~\cite{roy06}. 

Having known the drag and diffusion,  we
proceed to solve the FP equation. For this purpose
we require the initial parton distributions 
which is  parametrized as~\cite{mueller}:
\bea
f(p_T,p_z,t=t_i)\equiv \frac{dN}{d^2p_Tdy}|_{y=0}=
\frac{N_0}{(1+\frac{p_T}{p_0})^\alpha},
\eea
where $p_0$, $\alpha$ and $N_0$ are parameters. 
Solving the FP equation with the boundary conditions, 
$f(\vec{p},t)\rightarrow 0$ for $|p|\rightarrow \infty$,
we are ready to evaluate 
the nuclear suppression factor, $R_{AA}$ defined as \cite{dEnterria05},
\bea
R_{AA}(p_T) 
&=&\frac{\mathrm ``Hot~QCD~medium"}{\mathrm ``QCD~vacuum"}\nn\\
&=& \frac{\sum_a \int f_a({\bf p^{\prime}}, \tau_c)|_{p_T^{\prime} = p_T/z}
D_{a/\pi^0}(z,Q^2)dz}{\sum_a \int f_a({\bf p^{\prime}},\tau_i)
|_{p_T^{\prime} = p_T/z}, 
D_{a/\pi^0}(z,Q^2)dz}
\eea
\begin{figure}[*htb]
\epsfig{figure=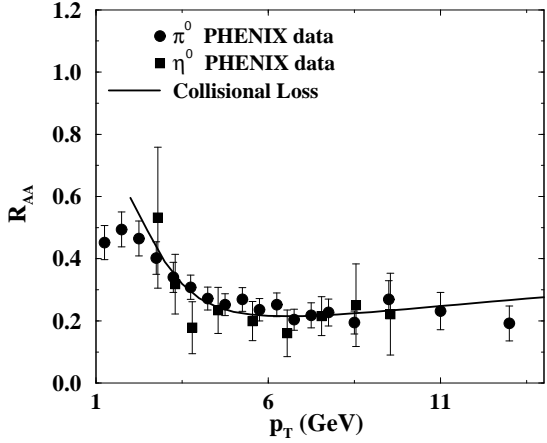, height= 6 cm}
\caption{
Nuclear suppression factor for pion. Experimental data are taken from
PHENIX collaboration 
\protect\cite{nuclex06} 
for Au + Au collisions
at $\sqrt{s}=200$ GeV. Solid  line indicates result from the
present calculation with collisional energy loss of the partons
propagating through the plasma 
before fragmenting into pions.
}
\label{fig1}
\end{figure}
where $f({\bf p^\prime},\tau_i)$ and $f({\bf p^\prime},\tau_c)$ 
denote the   parton distributions at proper time $\tau_i$ and $\tau_c$ 
respectively. Here $\tau_i$ is the initial time and $\tau_c$ is the time
when the system cools down to the transition 
temperature $T_c$ (=190 MeV)~\cite{katz}. The result for neutral pion
is shown in 
Fig.~\ref{fig1} which describes the PHENIX data\cite{nuclex06} for $Au+Au$ at
$\sqrt{s}=200$ GeV reasonably well. 


It should be noted  here that the $R_{AA}(p_T)$
with collisional loss has a tendency to increase
for higher $p_T$, indicating less importance of 
collisional loss at this domain, where the radiative
loss may become important. Therefore, a detailed
calculation with both collisional and radiative 
loss may be useful to delineate the importance
of individual mechanism.

\begin{figure}[*htb]
\epsfig{figure=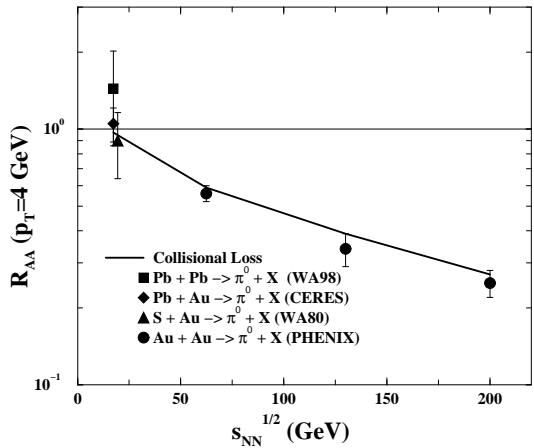, height= 6 cm}
\caption{ 
Excitation function of the nuclear modification factor for neutral pions
in central A+A reactions at a fixed $p_T=4$ GeV where 
only the elastic energy loss is considered. Experimental data 
are taken from~{\protect\cite{dEnterria05}}.
}
\label{fig2}
\end{figure}

To stress our point
further we also analyse the excitation function of the
nuclear suppression factor.
Results are shown 
in Fig.~\ref{fig2}. It is clear that
$R_{AA}(p_T)$ at $p_T=4$ GeV for various beam 
energies are found to be well described.
The values of  parton (quarks, anti-quarks and gluons) densities
($n_{g+q+\bar{q}}$)  of  the QCD medium  which describe
the data for various beam energies are shown in table I.

\begin{figure}[*htb]
\epsfig{figure=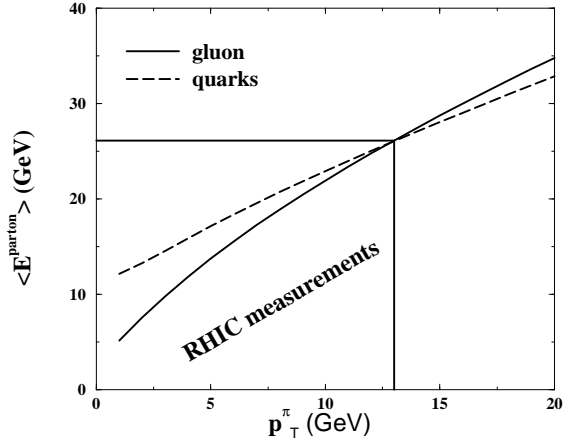, height= 6 cm}
\caption{ Average parton energy versus transverse momentum of pion for 
$\sqrt{s}=200$ GeV/A.}
\label{fig3}
\end{figure}

To pin down the relative importance of 
$2 \leftrightarrow 2$ $2\rightarrow 3$ processes, we  determine the average
energy of the parton which contribute to the measured 
$p_T$ window of the hadrons. 
To this end, the average fractional momentum
($\langle z\rangle$) of the fragmenting partons carried by the pion is
calculated using relevant parton distribution and fragmentation functions. 
For the former, we use CTEQ~\cite{cteq} including shadowing {\em via} EKS98 
parametrization\cite{eks}, while for the fragmentation function KKP 
parametrization is used \cite{kkp}.
The average energy of the parton, 
$\langle E^{\mathrm{parton}}\rangle$ is obtained by using the relation 
$<E^{parton}>=p_T^{\pi}/<z>$ for $y_{\pi} = 0$.
The results are shown in Fig.~\ref{fig3}. 
Our results are consistent with that of ref.\cite{dEnterria05} which
quotes $\langle z \rangle = p_{\mathrm {hadron}}/p_{\mathrm parton} 
\simeq 0.5-0.7$ for $p_{\mathrm {hadron}} \geq 4$ GeV at RHIC energies.

\begin{table}
\begin{tabular}{cccc}
\hline
$\sqrt{s_{NN}}$ &  $n_{g+q+\bar{q}}$ & $R_{AA}$ \\
\hline
\\
17.3  & 4  &  0.97\\  
62.4  & 15 &  0.59 \\
130   & 27 &  0.39 \\
200   & 49 &  0.27\\
\hline
\end{tabular}
\caption{The extracted initial parton densities and nuclear modification 
factors at various beam energies}
\end{table}

It might be recalled that at RHIC energies the nuclear modification 
factor $R_{AA}(p_T)$ has been measured in the pion transverse 
momenta range $p_T\sim 1-13$ GeV. 
Assuming that these pions are originated from the fragmenting 
partons, we ask the question, what is the average parton energy 
required to produce these pions? From Fig.~\ref{fig3}, it is 
clear that the maximum average parton energy required is about 26 GeV here. 

Now the next question is, what is the dominant
energy loss mechanism for partons with energy $\sim 26$ GeV or less?
We might compare this value with the determined $E_c$\footnote{Note
that  $E_c$ is defined to be the energy below which 
elastic loss dominates~\cite{abhee05}.} given in 
refs.~\cite{abhee05}(can also be read out from~\cite{djordjevic06})
 and note that at these energies
collisional loss cannot be neglected. For lower beam energy,  62.4 (130) 
AGeV the value of maximum average parton energy required to produce
a 13 GeV pion is 16 (22) GeV, where 
the collisional loss will definitely be more important. 
It is worthwhile to mention here that
this estimation of $E_c$ has some uncertainty as it depends on the length 
of the plasma, initial temperature, mean free path, dynamical screening 
mass etc. Those would affect both the mechanisms ({\it i.e.}
radiative and collisional) of energy loss. 
Our chosen parameter set is consistent with that of ref.\cite{zakharov,ewang} 
used to study the radiative energy loss.

In conclusion,
our investigations clearly suggest that in the measured
$p_T$ range of light hadrons at RHIC
collisional, rather than the radiative, is the dominant mechanism of 
jet quenching. This is in sharp contrast to all the previous analyses.
The determination of the critical energy($E_c$), however,
might change depending upon the detailed model of 
`jet quenching'. Inclusion of three body elastic channels for
heavy quark energy loss, which are considered in ref.\cite{ko06}, if 
applied for light flavours,  might even increase $E_c$, 
making our point stronger. $E_c$ will also increase if there exist
partonic bound state in the plasma due to 
ionization loss~\cite{shuryak}.
It should be mentioned that for the collisional energy loss we have
not included finite size effect which however is shown to be small 
\cite{djordjevic06}. In light of these new findings the theory of
jet tomography is expected to change considerably.

{\bf Acknowledgment} We are grateful to David d'Enterria for providing us 
the experimental data shown in  Fig. \ref{fig2}. We thank Sourav Sarkar
for useful discussions.

\end{document}